\documentclass{article}

\usepackage[utf8]{inputenc} 
\usepackage[T1]{fontenc}    
\usepackage{hyperref}       
\usepackage{url}            
\usepackage{booktabs}       
\usepackage{amsfonts}       
\usepackage{nicefrac}       
\usepackage{microtype}      
\usepackage{lipsum}		
\usepackage{graphicx}
\usepackage{doi}

\title{Reply to the "Comment on The Tully-Fisher law and dark matter effects derived via modified symmetries by I. Arraut"}

\author{ Ivan Arraut\\
\quad Institute of Science and Environment, University of Saint Joseph\\
  Estrada Marginal da Ilha Verde, 14-17, Macao, China\\
	\texttt{ivan.arraut@usj.edu.mo}
}

\hypersetup{
pdftitle={A template for the arxiv style},
pdfsubject={q-bio.NC, q-bio.QM},
pdfauthor={David S.~Hippocampus, Elias D.~Striatum},
pdfkeywords={First keyword, Second keyword, More},
}

\begin{document}
\maketitle

\begin{abstract}
\end{abstract}


\section{Introduction}

It has been claimed in \cite{1}, that the idea proposed in \cite{2} has certain mistakes based on arguments of energy conditions and others. Additionally, some of the key arguments of the paper are criticized. Here we demonstrate that the results obtained in \cite{2} are correct and that there is no violation of any energy condition. The statements claimed in \cite{1} are based on three things: 1). Misinterpretation of the metric solution. 2). Language issues related to the physical quantities obtained in \cite{1}, where the authors make wrong interpretations about certain results over the geometry proposed in \cite{2}. 3). Non-rigorous evaluations of the vacuum condition defined via the result over the Ricci tensor $R_{\mu\nu}=0$.

\section{The proposed metric}

The proposed metric in \cite{2}, is defined as

\begin{eqnarray}   \label{SDSmetric}
ds^2=-e^{2\alpha(r)}dt^2+e^{2\beta(r)}dr^2+\zeta rd\Omega^2,    
\end{eqnarray}
where $e^{2\alpha(r)}=e^{-2\beta(r)}=1-\frac{2GM}{r}$ and $\zeta$ is a parameter with the units of distance. This parameter is absorbed by the parameter $\gamma$ proposed in \cite{2} as the new invariant at galactic scales. Indeed, $\gamma^2=L^2/r$, but the author could perfectly use $\eta^2=\zeta L^2/r$ and the results would not change. Unfortunately, the authors in \cite{1} made a wrong statement saying that $r$ and $t$ are dimensionless in this case. All what we have done is to set a parameter $\zeta=1$ for simplicity, in the same way as sometimes we set the speed of light to one ($c=1$) for facilitating certain calculations. We have derived rigorously the solution (\ref{SDSmetric}) and we have demonstrated that it is asymptotically flat and that no energy condition is violated so far. In order to see this, we have to start from the most general form of the metric 

\begin{equation}   \label{2eq}
ds^2=-e^{2\alpha(r)}+e^{2\beta(r)}dr^2+e^{2\kappa(r)}r^2d\Omega^2.
\end{equation}
Here, we can redefine $\bar{r}\zeta=r^2e^{2\kappa(r)}$. After evaluating $d\bar{r}$, and introducing again inside eq. (\ref{2eq}); it is always possible to find a form of the metric with the angular part as in eq. (\ref{SDSmetric}), after considering $\left(1+r\frac{d\kappa}{dr}\right)^{-2}\frac{\zeta^2}{4r^2}e^{2\beta(r)-2\kappa(r)}\to e^{2\beta(r)}$, and $\bar{r}\to r$. These re-definitions are standard and are always used at the moment of finding solutions \cite{Carrol}. This part is trivial but still it does not guarantee that the solution (\ref{SDSmetric}) is correct. In order to verify that the solution (\ref{SDSmetric}) is the appropriate one, namely, a valid vacuum solution of the Einstein equations, we have to calculate the Ricci tensor $R_{\mu\nu}$ and verify the condition $R_{\mu \nu}=0$. For this purpose, we have to start from the ansatz metric in eq. (\ref{SDSmetric}), assuming that we do not know yet the functional dependence of $\alpha(r)$ and $\beta(r)$. In other words, we will find the solutions for this functions. After calculating the Riemann tensor and its contractions, we found that the non-zero components are 

\begin{eqnarray*}
R_{tt}^{new}=R_{tt}^{stand}-e^{2(\alpha-\beta)}\frac{\partial_r\alpha}{r},\;R_{rr}^{new}=R_{rr}^{stand}-\frac{1}{r}\partial_r\beta\nonumber\\
+\frac{1}{2r^2},\;R_{\theta\theta}^{new}=\frac{\zeta e^{-2\beta(r)}}{2}(\partial_r\beta-\partial_r\alpha)+1, R_{\phi\phi}=sin^2\theta R_{\theta\theta}.
\end{eqnarray*}
Here $R^{new}_{\mu\nu}$ is the Ricci tensor obtained under the metric (\ref{SDSmetric}), while $R_{\mu\nu}^{stand}$ is the same Ricci tensor component, but obtained under the standard spherically symmetric Schwarzschild case. Imposing the condition $R_{\mu\nu}=0$ over each component, gives us the right to write $e^{2(\beta(r)-\alpha(r))}R_{tt}+R_{rr}=0$. This condition is translated into $\bar{\alpha}(r)=-\bar{\beta}(r)$, with $\partial_r\bar{\alpha}(r)=\partial_r{\alpha}(r)+\frac{1}{4r}$. Finally, the condition $R_{\theta\theta}=0$, gives us a functional form $e^{2\bar{\alpha}(r)}= 1-\frac{A}{r}$, which is asymptotically flat. Then if we make a redefinition $\alpha(r)\to\bar{\alpha}(r)$ in the metric (\ref{SDSmetric}), which is always possible without violating any of the previous arguments, we obtain the desired result which is asymptotically flat, wthout violating any enrgy condition.   

\section{On the new symmetry condition}

The metric (\ref{SDSmetric}), together with the spatial Killing vector definition $K^\mu=(0, 0, 0, 1)$, gives us the one-form $K_\mu=\zeta r$. Then the invariant quantity is $K_\mu dx^{\mu}/d\lambda=\zeta rd\phi/d\lambda$, which under $\zeta=1$ corresponds to $\gamma$ \cite{2}. Although from the perspective of the metric (\ref{SDSmetric}), $\gamma$ is the angular momentum; we still continue using the word "angular momentum", for the equivalent quantity derived from the standard Schwarzschild metric, which appears through the invariant $L=r^2d\phi/d\lambda$. In this way, we define the relation $\gamma^2=L^2/r$ in \cite{2} and we interpret this quantity as our instruments would do it, namely, the square of the tangential velocity. Then there is no mistake in the arguments used in \cite{2} at this level.

\section{The baryonic Tully-Fisher law and MOND}

We must remark that the results obtained in \cite{2}, reproduce the Dark Matter effects and they are independent of MOND \cite{MOND}. The author in \cite{2} cited MOND when the baryonic Tully-Fisher law ($v_\phi^4=4GMa_0$) was mentioned. This relation appears when an acceleration scale $a_0$ is imposed and this is not exclusive to the case addressed in \cite{2}. Actually the Tully-Fisher law appears in the standard theory of gravity whenever the parameter $a_0=v_\phi^2/r$ (centripetal aceleration) is imposed. This acceleration scale is defined in the standard way (with respect to the Schwarzschild metric) and not with respect to the metric (\ref{SDSmetric}). In the same way as it was explained before, this is what generates the misunderstanding with respect to the statements claimed in \cite{1}. The author in \cite{2} tried to keep loyal to the standard definitions of acceleration, angular momentum and others; keeping in mind that these concepts have to be used inside the results obtained via the metric (\ref{SDSmetric}). Then for example, the standard equilibrium condition for a test particle moving around a source with a Schwarzschild metric is $r_{eq}\approx L^2/2GM$. Taking the standard definition of angular momentum (with respect to the standard Schwarzschild metric) $L=r(rd\phi/d\lambda)$), with $v_\phi=rd\phi/d\lambda$; and considering the acceleration parameter $a_0$, we get the baryonic version of the Tully-Fisher law \cite{2}. The same result is obtained from the equilibrium condition inside the solution proposed in \cite{2}, but taking into account that the invariant quantity is $\gamma$ and not $L$. What makes special the result obtained in \cite{2} is not the Tully-Fisher law given by $v_\phi^4=4GMa_0$. What is truly special is the fact that the proposed solution eliminates the problem of balancing the Newtonian and the centrifugal contributions of the potential. It also gives higher importance to the term coupling $GM$ with $L$. Through this term, the faster an object rotates, the larger will be the attraction to the source, something truly desirable. The potential obtained through the metric (\ref{SDSmetric}), and given as \cite{2}

\begin{equation}   \label{Potential1}
V_1(r)=-\frac{GM}{r}+\frac{\gamma^2}{2r}-\frac{GM\gamma^2}{r^2}.    
\end{equation}
Then the condition $\gamma^2\to2GM$, defines a transition regime. If $\gamma^2>2GM$, then the centrifugal term is dominant, while when $\gamma^2<2GM$, gravity is absolutely dominant. From this perspective, we can perceive the value $\gamma^2=2GM$ as a transition scale where there is an exact cancellation of the Newtonian contribution and the centrifugal one. The galaxy rotation curves can be reproduced if we locate the test particle at different distances $r$, noticing that the location does not affect the relation between the first two terms in the potential (\ref{Potential1}). The most important feature of the idea proposed in \cite{2}, is the irrelevance of the distance with respect to the source, at the moment of comparing the centrifugal contribution and the Newtonian contribution inside the potential (\ref{Potential1}). There is an additional (enhanced) contribution coming from the third term on the same potential, which in any case also contributes with an attractive effect increasing with the velocity of rotation of the test particle.

\section{Conclusions}

We have demonstrated that the results of the paper \cite{2} are correct. There is no violation of energy conditions and the solution is asymptotically flat after doing the right calculations. For the author, it would be really interesting to find some violations of the energy conditions mentioned in \cite{1}, because it would mean some possible connections between the Dark Matter effects and Dark Energy. Unfortunately, the author verified that such violations, claimed in \cite{1}, did not emerge. Finally, the authors in \cite{1}, did a wrong interpretation of the results proposed in \cite{2} because they did not keep the standard definitions of certain physical quantities when the worked over the metric (\ref{SDSmetric}). Finally, the baryonic version of the Tully-Fisher law appears whenever an acceleration scale is imposed. We must remark that the results obtained in \cite{2} do not correspond to the theory of MOND. They rather reproduce some key features of this theory, but derived from fundamental principles. The author of \cite{2}, recognizes a typos mistake after eq. (22) in the original paper, where  the right result should be $M_{eff}=M\left(1+\frac{\gamma^2}{r}\right)$. A longer version of the results illustrated in \cite{2}, explaining in detail all the calculations is in process.

\end{document}